\documentclass[11pt,a4paper,useAMS]{emulateapj}
\usepackage{amsmath}
\usepackage{epsfig}
\usepackage{natbib}
\usepackage{xcolor}
\usepackage{epsfig}
\usepackage{graphicx}
\usepackage{pifont}
\usepackage{xcolor}
\usepackage{hyperref}

\shorttitle{LITTLE RED DOTS IN X-RAYS}
\shortauthors{ANANNA ET AL.}

\begin{document}

\title{X-ray View of Little Red Dots: \\ Do They Host Supermassive Black Holes?\vspace{-10ex}}
\author{Tonima Tasnim Ananna\altaffilmark{1}, \'Akos Bogd\'an\altaffilmark{2}, Orsolya E. Kov\'acs\altaffilmark{3}, Priyamvada Natarajan\altaffilmark{4,5,6}, Ryan C. Hickox\altaffilmark{7}}
\affil{\altaffilmark{1} Department of Physics and Astronomy, Wayne State University, Detroit, MI 48202, USA}
\affil{\altaffilmark{2}Center for Astrophysics \ding{120} Harvard \& Smithsonian, 60 Garden Street, Cambridge, MA 02138, USA}
\affil{\altaffilmark{3} Department of Theoretical Physics and Astrophysics, Faculty of Science, Masaryk University, Kotl\'a\v{r}sk\'a 2, Brno, 611 37, Czech Republic}
\affil{\altaffilmark{4} Department of Astronomy, Yale University, New Haven, CT 06511, USA}
\affil{\altaffilmark{5} Department of Physics, Yale University, New Haven, CT 06520, USA}
\affil{\altaffilmark{6} Black Hole Initiative, Harvard University, 20 Garden Street, Cambridge, MA 02138, USA}
\affil{\altaffilmark{7} Department of Physics and Astronomy, Dartmouth College, 6127 Wilder Laboratory, Hanover, NH 03755, USA}

\begin{abstract}
The discovery of Little Red Dots (LRDs) -- a population of compact, high-redshift, dust-reddened galaxies -- is one of the most surprising results from \textit{JWST}. However, the nature of LRDs is still debated: {does the near-infrared emission originate from accreting supermassive black holes (SMBHs), or intense star formation?}
In this work, we utilize ultra-deep \textit{Chandra} observations and study LRDs residing behind the lensing galaxy cluster, Abell~2744. We probe the X-ray emission from individual galaxies but find that they remain undetected and provide SMBH mass upper limits of $\lesssim(1.5-16)\times10^{6}~\rm{M_{\odot}}$ assuming Eddington limited accretion. To increase the signal-to-noise ratios, we conduct a stacking analysis of the full sample with a total lensed exposure time of $\approx87$~Ms. We also bin the galaxies based on their stellar mass, lensing magnification, and detected broad-line H$\alpha$ emission. For the LRDs exhibiting broad-line H$\alpha$ emission, there is a hint of a stacked signal ($\sim2.6\sigma$), corresponding to a SMBH mass of $\sim3.2\times10^{6}~\rm{M_{\odot}}$. Assuming unobscured, Eddington-limited accretion, this BH mass is at least 1.5 orders of magnitude lower than that inferred from virial mass estimates using \textit{JWST} spectra. Given galaxy-dominated stellar mass estimates, our results imply that LRDs do not host over-massive SMBHs and/or accrete at a few percent of their Eddington limit. {However, alternative stellar mass estimates may still support that LRDs host over-massive BHs.} The significant discrepancy between the \textit{JWST} and \textit{Chandra} data hints that the scaling relations used to infer the SMBH mass from the H$\alpha$ line and virial relations may not be applicable for high-redshift LRDs.
\end{abstract}


\section{Introduction}
\label{sec:intro}

Since its launch, \textit{JWST} has revolutionized the picture of the early universe on multiple forefronts. In particular, \textit{JWST} has discovered an unexpectedly large sample of high-redshift galaxies in multiple wide fields and behind lensing galaxy clusters \citep[e.g.][]{robertson22,castellano22a,castellano22b,atek23}. \textit{JWST} has also brought into view high-redshift ($z>8$) accreting SMBHs that shed new light on our understanding of early seeding and growth \citep{bogdan24,natarajan24,kokorev23,kovacs24,maiolino24}. In addition, thanks to its near-infrared sensitivity, \textit{JWST} unveiled a previously unknown, dust-reddened, galaxy population at high ($3<z<9$) redshifts \citep{labbe23a,labbe23b,matthee23,akins23,barro24}. These galaxies, dubbed as Little Red Dots (LRDs), are red, which suggests heavy obscuration, are compact ($r_{\rm e}\sim50 \ \rm{pc}$), and many of them have high inferred stellar masses ($M_{\rm \star} \gtrsim 10^{10} \ \rm{M_{\odot}}$).

The discovery of LRDs has triggered a debate about their nature. The emission in LRDs could be either dominated by flux from active galactic nuclei (AGN) or from a population of young stars associated with vigorous star formation. Alternatively, LRDs could be a heterogeneous galaxy population, in which both AGN and star formation contribute to their fluxes to different degrees. To constrain the origin of LRDs, several systems were followed up with \textit{JWST}, revealing evidence for both of these possibilities. For example, \citet{kocevski23} studied the \textit{JWST} spectrum of two LRDs and found broad H$\alpha$ emission, indicating the presence of SMBHs with $M_{\rm BH } \sim 10^{7}\ \rm{M_{\odot}}$. Similar spectroscopic follow-up studies were performed for several other LRDs, which also suggest that they could host SMBHs with  $M_{\rm BH } \sim 10^{7}-10^{9}\ \rm{M_{\odot}}$ \citep{greene23,killi23,matthee23}. The high SMBH masses imply that the black hole-to-stellar mass ratio of LRDs is much higher than that in the local universe \citep{magorrian98,haring04,kormendy13,reines15} or for AGN observed up to $z\sim2.5$ \citep{suh20}. This result is hardly unexpected: previous \textit{Chandra}/\textit{JWST} studies found several high-redshift galaxies that host extremely massive SMBHs with some of them reaching black hole-to-stellar mass ratios of $\sim10\%-100\%$  \citep[e.g.][]{bogdan24,goulding23,ubler23,harikane23,kovacs24}. However, other studies, based on \textit{JWST} MIRI imaging, suggest that the emission in several LRDs is dominated by young stellar populations. Specifically, modeling the spectral energy distribution of 31 LRDs indicates that their characteristics can be described by intense and compact starbursts and young stellar ages, where the energy output is dominated by the emission from OB stars \citep{perez24}. An alternative interpretation using panchormatic MIRI photometry and ALMA data is that these are post-starburst galaxies with moderate dust level \citep{williams2023}.

\begin{figure*}[!htp]
  \begin{center}
    \leavevmode
      \epsfxsize=1\textwidth \epsfbox{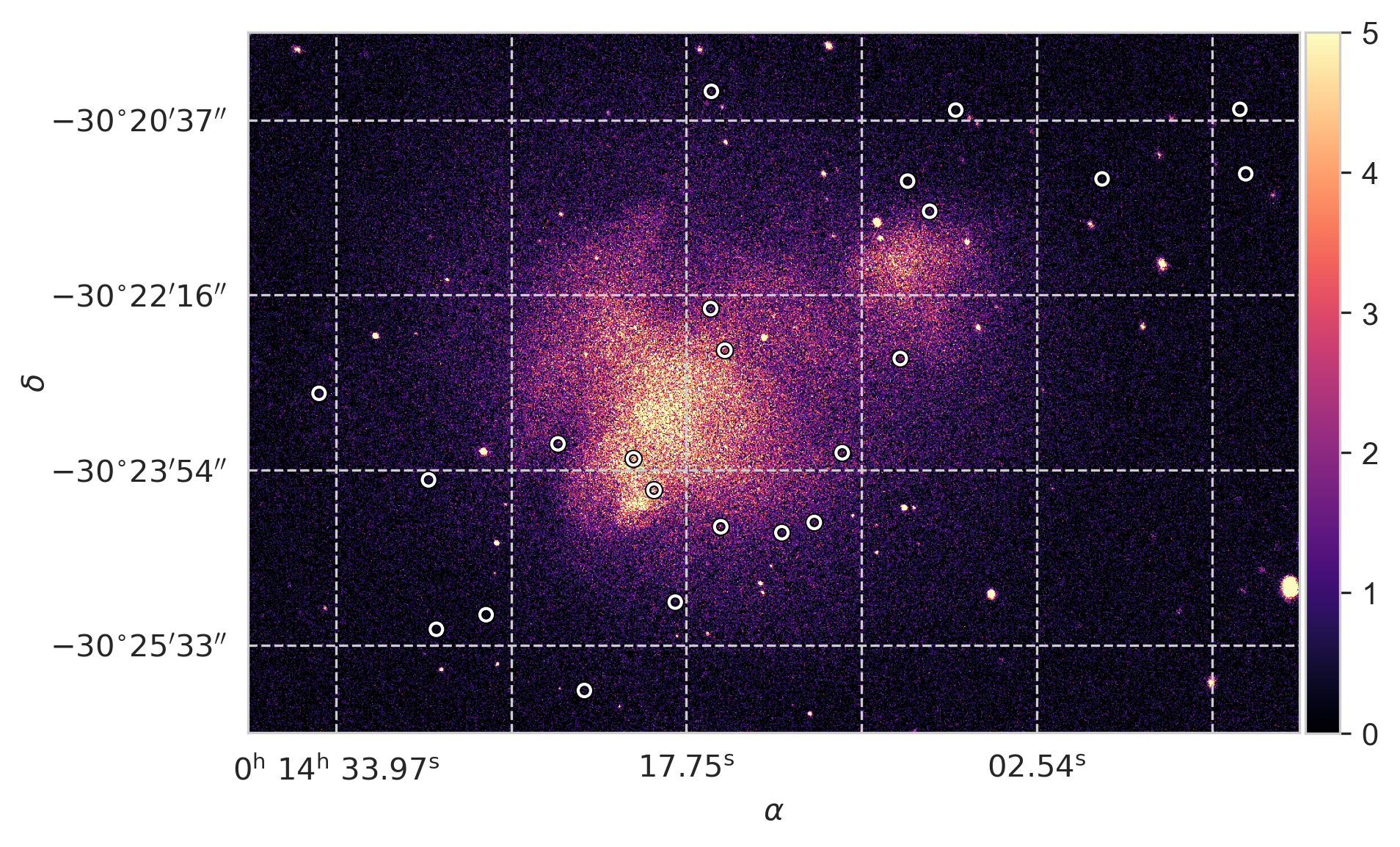}
      \vspace{0.0cm}
      \caption{Merged \textit{Chandra} ACIS-I image of the lensing cluster, Abell~2744, in the $2-7$~keV band. The total exposure time of the merged data is $\approx2.07$~Ms. We focus on a sample of 23 LRDs (see Table~\ref{tab:sample}) in the redshift range of $z=3-8.5$ that were identified by \textit{JWST} NIRCAM data \citep{labbe23b}. The location of the galaxies is highlighted with small white circles. The \textit{Chandra} X-ray data provides data for all galaxies.} 
     \label{fig:chandra}
  \end{center}
\end{figure*}

Although \textit{JWST} carried out several follow-up observations with the NIRSpec and MIRI instruments to better understand the nature of these galaxies, interpreting the near-infrared data is often not trivial because AGN and star-forming processes could produce very similar spectral signatures. X-ray data, in contrast, can provide important clues about the nature of LRDs for several reasons. First, the X-ray emission from dust-obscured AGN emission is less affected by obscuration. Although a larger fraction of AGN are obscured at higher redshifts, the shift to higher rest-frame X-ray energies allows the detection of heavily obscured AGN  \citep[e.g.][]{bogdan22,bogdan24,kovacs24}. Second, the level of expected X-ray emission from AGN and star-forming processes is significantly different with the latter being several orders of magnitude fainter. Therefore, given the typical exposure times of deep X-ray observations and the sensitivity of present-day X-ray telescopes, emission from star-forming processes will likely remain undetected, while accreting SMBHs could be detected even at large cosmic distances. Finally, the ultra-deep \textit{Chandra} observations of Abell~2744, combined with the lensing magnification of the galaxy cluster, allow the detection of accreting SMBHs with $10^7-10^8 \ \rm{M_{\odot}}$ even at $z\approx10$ \citep{bogdan24,kovacs24}. These SMBH masses are within the range of those predicted by \textit{JWST} spectroscopic studies of LRDs \citep{greene23,matthee23}. Therefore, the \textit{Chandra} X-ray observations of Abell~2744 are ideally suited to independently probe AGN in LRDs.

Despite the advantages of the X-ray waveband, searching for AGN in high-redshift galaxies remains challenging due to the relatively small collecting area of present-day X-ray observatories. To alleviate this problem, we utilize ultra-deep X-ray observations and boost the sensitivity of X-ray telescopes by gravitational lensing. 
In this work, we rely on \textit{Chandra} X-ray observations of the lensing galaxy cluster, Abell~2744, to probe the population of \textit{JWST}-detected LRDs behind this cluster \citep{labbe23b}. Our goal is to search for X-ray emission from AGN in individual LRDs and to also probe the collective X-ray emission from this galaxy population by stacking their X-ray photons. The deep \textit{Chandra} observations, combined with the lensing magnification, allow us to investigate whether LRDs host massive BHs, constrain the black hole-to-galaxy mass ratio, and assess the sample-averaged X-ray properties of LRDs. 

Our paper is structured as follows. Section \ref{sec:sample} describes the analyzed sample of LRDs. The analysis of the \textit{Chandra} X-ray data is outlined in Section \ref{sec:data}. The results on the X-ray emission from individual galaxies and on the collective X-ray emission from LRDs are presented in Section \ref{sec:results}. We conclude and place our results in a broader context in Section \ref{sec:discussion}. In this paper, the following cosmological parameters are assumed: $H_0=71 \ \rm{km \ s^{-1} \ Mpc^{-1}}$, $ \Omega_M=0.27$, and $\Omega_{\Lambda}=0.73$.

\begin{table*}[!t]
\caption{\textcolor{black}{Analyzed sample of Little Red Dots.}}
\begin{minipage}{18cm}

\renewcommand{\arraystretch}{1.3}
\centering
\begin{tabular}{lcccccccccc}
\hline
ID & RA & Dec & z & $\mu$ & $\log (M_{\rm \star})$ & $\log (L_{\rm 2-10 keV})$ & $\log (L_{\rm bol})$ & $ \log (M_{\rm BH})$ & $M_{\rm BH}/M_{\rm \star}$  \\
(1) & (2) & (3) & (4) & (5) & (6) & (7) & (8) & (9) & (10) \\
\hline
571$^\ddagger$	& 	3.592423	& 	$-$30.432824 	& 	6.74$^{\rm 	\star}$ 	& 	1.69	&  	9.0$\pm$0.1  	& 	$<43.68$	& 	$<44.91$	& 	$<6.81$	& 	$<6.48 \times 10^{-3}$		\\
1967	& 	3.619200	& 	$-$30.423270 	& 	5.84$^{\rm 	\star}$ 	& 	1.63	&  	10.9$\pm$0.1  	& 	$<43.39$	& 	$<44.61$	& 	$<6.51$	& 	$<4.03 \times 10^{-5}$		 \\
2476	& 	3.610205	& 	$-$30.421001 	& 	4.56	        	& 	1.92	&  	10.0$\pm$0.1  	& 	$<43.13$	& 	$<44.34$	& 	$<6.24$	& 	$<1.74 \times 10^{-4}$		\\
2940$^\ddagger$	& 	3.575989	& 	$-$30.419030 	& 	4.47	        	& 	1.81	&  	9.0$\pm$0.1   	& 	$<43.35$	& 	$<44.57$	& 	$<6.47$	& 	$<2.92 \times 10^{-3}$	 \\
5957$^\ddagger$	& 	3.556703	& 	$-$30.408192 	& 	3.52	        	& 	1.57	&  	10.4$\pm$0.1  	& 	$<43.25$	& 	$<44.46$	& 	$<6.36$	& 	$<9.05 \times 10^{-5}$		 \\
6151	& 	3.567758	& 	$-$30.407272 	& 	4.45	        	& 	1.86	&  	8.7$\pm$0.1   	& 	$<43.46$	& 	$<44.68$	& 	$<6.58$	& 	$<7.66 \times 10^{-3}$	\\ 
6430$^\ddagger$	& 	3.550837	& 	$-$30.406598 	& 	5.05$^{\rm 	\star}$ 	& 	1.54	&  	10.1$\pm$0.1  	& 	$<43.57$	& 	$<44.80$	& 	$<6.70$	& 	$<3.96 \times 10^{-4}$		 \\
8296$^{\dagger \ddagger}$	& 	3.579829	& 	$-$30.401569 	& 	7.04$^{\rm 	\star}$ 	& 	6.94	&  	8.8$\pm$0.1  	& 	$<43.47$	& 	$<44.69$	& 	$<6.59$	& 	$<6.17 \times 10^{-3}$		\\
8798$^\ddagger$	& 	3.620605	& 	$-$30.399950 	& 	6.34$^{\rm 	\star}$ 	& 	1.58	&  	9.6$\pm$0.1  	& 	$<43.49$	& 	$<44.71$	& 	$<6.61$	& 	$<1.03 \times 10^{-3}$	 \\
9992$^{\dagger \ddagger}$	& 	3.583536	& 	$-$30.396676 	& 	7.04$^{\rm 	\star}$ 	& 	8.55	&  	8.9$\pm$0.1  	& 	$<43.38$	& 	$<44.59$	& 	$<6.49$	& 	$<3.92 \times 10^{-3}$		\\
10148$^\ddagger$	& 	3.545794	& 	$-$30.395724 	& 	4.96	        	& 	1.69	&  	9.8$\pm$0.1  	& 	$<43.58$	& 	$<44.81$	& 	$<6.71$	& 	$<8.05 \times 10^{-4}$		\\
10712$^{\dagger \ddagger}$	& 	3.597201	& 	$-$30.394328 	& 	7.04$^{\rm 	\star}$ 	& 	4.12	&  	9.1$\pm$0.1  	& 	$<43.68$	& 	$<44.92$	& 	$<6.82$	& 	$<5.23 \times 10^{-3}$		 \\
13556$^\ddagger$	& 	3.640410	& 	$-$30.386436 	& 	8.50$^{\rm 	\star}$ 	& 	1.32	&  	9.0$\pm$0.1   	& 	$<44.03$	& 	$<45.30$	& 	$<7.20$	& 	$<1.58 \times 10^{-2}$		\\
15798	& 	3.535308	& 	$-$30.381010 	& 	6.44	        	& 	2.54	&  	8.0$\pm$0.2   	& 	$<43.74$	& 	$<44.98$	& 	$<6.88$	& 	$<7.63 \times 10^{-2}$		 \\
16561$^\ddagger$	& 	3.567022	& 	$-$30.379719 	& 	6.34	        	& 	3.56	&  	8.9$\pm$0.1   	& 	$<43.63$	& 	$<44.86$	& 	$<6.76$	& 	$<7.32 \times 10^{-3}$		\\
20080$^\ddagger$	& 	3.569595	& 	$-$30.373224 	& 	7.04$^{\rm 	\star}$ 	& 	2.71	&  	9.2$\pm$0.1  	& 	$<43.81$	& 	$<45.05$	& 	$<6.95$	& 	$<5.65 \times 10^{-3}$		\\
28343$^\ddagger$	& 	3.530008	& 	$-$30.358013 	& 	4.96$^{\rm 	\star}$  	& 	1.87	&  	9.8$\pm$0.1   	& 	$<43.71$	& 	$<44.94$	& 	$<6.84$	& 	$<1.11 \times 10^{-3}$		  \\
30782$^\ddagger$	& 	3.533997	& 	$-$30.353311 	& 	6.76$^{\rm 	\star}$ 	& 	1.75	&  	9.4$\pm$0.1  	& 	$<43.77$ & 	$<45.01$	& 	$<6.91$	& 	$<3.26 \times 10^{-3}$		  \\
31142	& 	3.498841	& 	$-$30.352945 	& 	4.81	        	& 	1.00	&  	10.2$\pm$0.1  	& 	$<43.71$	& 	$<44.95$	& 	$<6.85$	& 	$<4.46 \times 10^{-4}$		  \\
31298	& 	3.472870	& 	$-$30.352132 	& 	4.82	        	& 	1.00	&  	11.0$\pm$0.2  	& 	$<43.76$	& 	$<45.01$	& 	$<6.91$	& 	$<8.07 \times 10^{-5}$		 \\
35771	& 	3.525296	& 	$-$30.342213 	& 	4.30	        	& 	1.44	&  	10.1$\pm$0.2  	& 	$<43.45$	& 	$<44.67$	& 	$<6.57$	& 	$<2.93 \times 10^{-4}$		\\
35819	& 	3.473941	& 	$-$30.342061 	& 	4.59	        	& 	1.00	&  	8.9$\pm$0.1   	& 	$<43.50$	& 	$<44.72$	& 	$<6.62$	& 	$<5.27 \times 10^{-3}$		\\
37108	& 	3.569464	& 	$-$30.339305 	& 	3.01	        	& 	1.53	&  	9.1$\pm$0.1   	& 	$<43.08$	& 	$<44.28$	& 	$<6.18$	& 	$<1.53 \times 10^{-3}$		\\
 \hline
\label{tab:sample}
\end{tabular} 
\end{minipage}
$^{\dagger}$  Triple-lensed quasar discussed in \citet{furtak23}. Only ID8296 was included in the stacking analysis.\\
$^\ddagger$ AGN subset including 12 galaxies as identified by \citet{labbe23b}.\\
Columns are as follows: (1) UNCOVER ID;  (2) and (3): R.A. and Dec. (4) Redshift of the galaxies, where $z^{\rm \star}$ are spectroscopic redshifts adopted from \citet{greene23}); (5) Lensing magnification; (6) Stellar mass adopted from \citet{labbe23b} from SED fitting using ALMA data; (7) $2-10$~keV band X-ray luminosity; (8) Bolometric luminosity inferred from column (7) and the X-ray-to-bolometric correction of \citet{duras20}; (9) BH mass inferred from the column (9) assuming Eddington limited accretion; and (10) BH-to-galaxy stellar mass ratio. 
\end{table*}

\section{Sample of Little Red Dots}
\label{sec:sample}

We study a sample of high-redshift LRDs in the Abell~2744 field that were detected by \textit{JWST} \citep{labbe23b}. From the parent sample of 35 red and compact galaxies, \citet{labbe23b} identified 26 objects that exhibit red colors ($\rm{F200W-F444W} = 1 - 4$), a compact nature with a median size of $r_{\rm eff} \sim 50$~pc, and a light profile that is dominated by a central point source-like component. Follow-up spectroscopic studies of these 26 systems demonstrated that three of them are nearby brown dwarfs\citep{burgasser24} and three of them are associated with a triply imaged lensed galaxy at $z=7.045$ \citep{furtak23}. This leaves us with 21 individual LRDs that represent our main sample. Within this sample, \citet{labbe23b} identified an AGN subset of 12 galaxies whose broadband SED modeling required an AGN component. A sub-sample of LRDs was followed up by \textit{JWST} spectroscopy \citep{greene23}, which indicated that nine LRDs in the Abell~2744 field host AGN with masses of $M_{\rm BH} = (0.2-8)\times10^8 \ \rm{M_{\odot}}$.  

The redshifts of the 21 individual LRDs are in the range of $z=3.0-8.5$ with a median of $z=4.96$. Of the 21 galaxies, 11 have precise spectroscopic redshifts \citep{greene23,burgasser24,furtak23}. The stellar masses of the galaxies span three orders of magnitude and are in the range of $M_{\star} =10^8- 10^{11} \ \rm{M_{\odot}}$ with a median of $M_{\star} =2.5\times10^9 \ \rm{M_{\odot}}$. Thanks to the lensing magnification of Abell~2744, the light from these galaxies is amplified by factors of $\mu = 1-8.55$ with a median of $\mu = 1.69$. The main properties of the galaxies are listed in Table \ref{tab:sample} and for details we refer to previous works \citep{labbe23b,greene23}.

\section{Data analysis}
\label{sec:data}
\subsection{The Chandra data}
\label{sec:chandra}
We analyzed 101 \textit{Chandra} ACIS-I observations of Abell~2744. Each of the observations was centered on the core of Abell~2744. Thanks to the large field of view of the ACIS-I array, the X-ray data covers all LRDs that were identified by \textit{JWST}. The total exposure time of the \textit{Chandra} data is $2.07$~Ms (for details see Table \ref{tab:data}). The data was processed with CIAO (version 4.15) and CALDB version 4.11 \citep{fruscione06}. 

The main steps of the data analysis are identical to those outlined in \citet{kovacs24}. We first reprocessed all observations with the \textsc{chandra\_repro} tool, which assures that uniform calibration is applied to all observations. When running this task, we applied the \textsc{vfaint} filtering. The high background periods were excluded, which decreased the exposure time by $\approx3\%$. For each observation, we generated exposure maps assuming a power law model with a slope of $\Gamma= 1.9$ and Galactic column density ($N_{\rm H} = 1.35\times10^{20} \ \rm{cm^{-2}}$) \citep{column}.

\begin{figure*}[!htp]
  \begin{center}
    \leavevmode
      \epsfxsize=0.495\textwidth \epsfbox{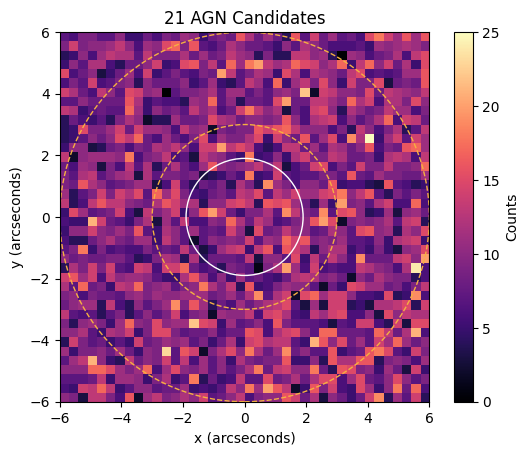}
      \epsfxsize=0.495\textwidth \epsfbox{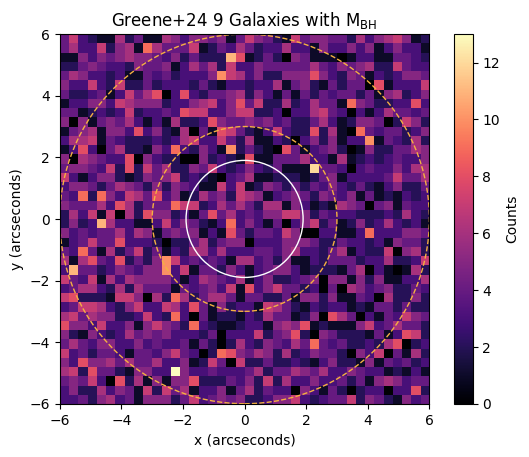}
      \epsfxsize=0.495\textwidth \epsfbox{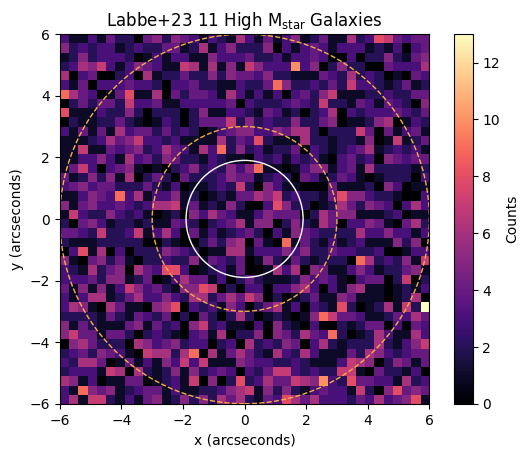}
      \epsfxsize=0.495\textwidth \epsfbox{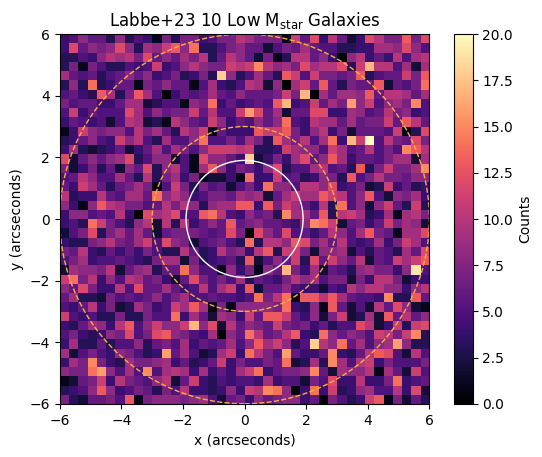}
      \vspace{0.0cm}
      \caption{Stacked $2-7$~keV band \textit{Chandra} X-ray images of LRD galaxies using different samples. The solid circular region represents a $1.9\arcsec$ region and the dashed annulus shows a $3\arcsec-6\arcsec$ annulus. The regions shown here are only for illustration purposes, the actual source and background regions used in our analysis were calculated using the local PSF for each observation (Section~\ref{sec:stackfast} and Appendix A). The stacked samples are as follows. \textit{Top left:} The 21 individual LRDs in our sample; \textit{Top right:} 9 LRDs from the \citet{greene23} sample with broad-line H$\alpha$ emission and an inferred SMBH mass. We obtain a tentative $2.6\sigma$ detection for this subset.; \textit{Bottom left:} 11 massive galaxies with stellar mass $\geq2.5\times10^{9} \ \rm{M_{\odot}}$; \textit{Bottom right:} 10 low-mass galaxies with stellar masses $<2.5\times10^{9} \ \rm{M_{\odot}}$.}  
     \label{fig:stacks}
  \end{center}
\end{figure*}

\subsection{Analysis with \textsc{StackFast}}
\label{sec:stackfast}
To analyze the individual observations and stack the X-ray photons associated with LRDs, we used the \textsc{StackFast} package \citep{ananna2023}. \textsc{StackFast} is a publicly available software that takes in positional data of sources, locates all the observations where these sources appear and stacks photons based on user-specified aperture and background sizes. The input aperture and background sizes are multiples of $90\%$ energy encircling radius around each source (henceforth, $R_{\rm 90}$). This radius depends on the distance of the source from the pointing center of the observation on which it falls. The $R_{\rm 90}$ for \textit{Chandra} is calculated as follows\footnote{Further details are provided in the \textit{Chandra} \href{https://cxc.harvard.edu/proposer/POG/html/chap4.html\#tth_sEc4.2.3}{proposer's guide}.}:

\begin{equation}\label{eqn:r90}
R_{\rm 90}=1.5+9 \times \left(\frac{{\rm Angular~Distance}}{10}\right)^{1.9} 
\end{equation}

For the analysis, the source apertures were fixed at $R_{\rm 90}$ and the background annuli have radii of $ (1.3-2)R_{\rm 90}$. The background radius is small because of the substantial variation associated with the ICM of Abell~2744, implying that a more localized background will produce accurate estimates of background/foreground values.

\begin{table*}[t]
\caption{\textcolor{black}{Upper limits on the SMBH masses of stacked samples}}
\begin{minipage}{18cm}
\renewcommand{\arraystretch}{1.3}
\centering
\begin{tabular}{cccccccccccc}
\hline
Bin & \#  &  $ z $ & $ \log (M_{\rm \star}) $ & $\log (L_{\rm 2-10 keV})$ & $\log (L_{\rm bol})$ & $ \log (M_{\rm BH})$ & $M_{\rm BH}/M_{\rm \star}$  \\
(1) & (2) & (3) & (4) & (5) & (6)  & (7)  & (8) \\
\hline

All 	&  21 &	4.96 &	    9.4     		& 	$<42.91$	& 	$<44.11$	& 	$<6.01$	& 	$<4.07\times 10^{-4}$	\\
AGN subset & 12 &	6.34 &	  9.3    		& 	$<43.28$	& 	$<44.49$	& 	$<6.39$	& 	$<1.23\times 10^{-3}$ \\

$M_{\rm \star} < $ $2.5\times10^{9}$	& 10 & 	6.44 &	        	9.0 	& 	$<43.32$	& 	$<44.54$	& 	$<6.44$	& 	$<2.75 \times 10^{-3}$	\\
$M_{\rm \star} \geq $ $2.5\times10^{9}$	&  11 & 	4.89 &	        10.1 		& 	$<43.08$	& 	$<44.28$	& 	$<6.18$	& 	$<1.20 \times 10^{-4}$	\\

$\mu < $ $1.69$	& 10 & 	4.81 &	10.1   		& 	$<43.07$	& 	$<44.27$	& 	$<6.17$	& 	$<1.17 \times 10^{-4}$	\\
$\mu \geq $ $1.69$	& 11 & 	6.34 &	    $9.0 $     		& 	$<43.29$	& 	$<44.50$	& 	$<6.40$	& 	$<2.50 \times 10^{-3}$	\\
Broad H$\alpha$ line & 9 & 6.26 & 9.5 & $43.38^{+0.14}_{-0.21}$ & $ 44.60^{+0.14}_{-0.21}$ & $ 6.50^{+0.14}_{-0.21}$ & $1.00 \times 10^{-3}$ \\
 \hline
\label{tab:limits}
\end{tabular}
\end{minipage}
Columns are as follows: (1) Binning method of the LRDs for the stacking analysis; (2) Number of galaxies in the stack; (3) and (4) Median redshift and stellar mass of the galaxies in the bin, respectively; (5) $2-10$~keV band X-ray luminosity; (6) Bolometric luminosity inferred from column (5) and the X-ray-to-bolometric correction of \citet{duras20}; (6) SMBH mass inferred from the column (6) assuming Eddington limited accretion; and (8) Black hole-to-galaxy stellar mass ratio. 
\end{table*}

\textsc{StackFast} also accounts for sudden fluctuations or the presence of other sources in the background region by dividing up the background annulus into twelve sectors of equal size and excluding regions with fewer than two counts and regions with counts 2$\sigma$ above the median of the twelve sectors. The exposure time is calculated by taking vignetting, bad pixels, and any areas within the extraction region that fall outside the ACIS-I detectors. In addition, \textsc{StackFast} takes a list of positions of known resolved sources within Abell~2744, and for each source-observation pair, if a source of interest (e.g., an LRD) fell within $1.5 \times R_{\rm 90}$ of a previously known X-ray source, that source-observation pair was excluded from our analysis.

\subsection{Stellar Mass Estimates}

In this work, the stellar mass estimates are taken from \citet{labbe23b}. Stellar mass estimates in \citet{labbe23b} are calculated using SED fitting, and the reported stellar masses are a result of galaxy-only templates fitted to the data. In \S~3.3 of \citet{labbe23b}, it is stated that sources that have ALMA data are better fitted with a dominant AGN component. Several other related works have pointed out \citet{akins2023,barro2024,wang2024} that including an AGN component can lead to stellar mass being lowered by 1-2 magnitudes. A galaxy+AGN template fit can therefore lower the stellar mass estimates used in this work, and will have implications for black hole mass to stellar mass ratio, which is further discussed in the \S~\ref{sec:discussion}.

\section{Results}
\label{sec:results}
\subsection{Individual galaxies}
\label{sec:individual}

The LRDs in our sample were identified as potential AGN host galaxies \citep{labbe23b,greene23}. Since SMBHs with masses of $10^7-10^8 \ \rm{M_{\odot}}$ accreting at their Eddington limit could be detectable in the Abell~2744 field with the deep \textit{Chandra} observations, we first carry out X-ray photometry at the position of individual LRDs. 

To perform the photometry, we utilize the coordinates of the galaxies as determined by \textit{JWST} and rely on the \textit{Chandra} ACIS-I images in the $0.5-2$~keV, $2-7$~keV, and $0.5-7$~keV bands and use the \textsc{StackFast} as described in Section \ref{sec:data}. None of the LRDs are detected at a significance $\geq2\sigma$ in the studied energy ranges. In the absence of detections, we place upper limits on the X-ray luminosity of these galaxies and derive upper limits on their SMBH masses, which are presented in Table \ref{tab:sample} and further discussed in Section \ref{sec:bh_mass}.

\subsection{Stacking the sample of LRDs}
\label{sec:stacking}

Because the individual LRDs remain undetected, we boost the sensitivity of the \textit{Chandra} data by co-adding the X-ray photons from the individual galaxies. To perform the stacking analysis, we utilize the X-ray data of the 21 individual LRDs as discussed in Section \ref{sec:stackfast}. We note that the triply lensed galaxy is included only once in our sample (ID~8296), which represents the brightest \textit{JWST} detection.  

The top left panel of Figure~\ref{fig:stacks} shows the stacked image of the 21 individual LRDs in the $2-7$~keV band and reveals a non-detection. We also report non-detections in the $0.5-2$~keV and $0.5-7$~keV bands. We note that the total stacked, lensed exposure time of the \textit{Chandra} X-ray data is $\approx8.7\times10^7$~s. Given these non-detections, we can place robust upper limits on the mean X-ray luminosity and SMBH mass of the LRDs, which are presented in Table \ref{tab:limits} and in Section \ref{sec:bh_mass}.  

While the full sample of LRDs does not reveal a statistically significant detection, we split our sample based on their stellar mass and lensing magnification to obtain potential detections. In the low-redshift universe, the mass of SMBHs correlates with the stellar mass of galaxies/bulges \citep[e.g.][]{magorrian98,haring04,kormendy13,reines13}. Assuming that a similar correlation exists at high redshifts, and more massive LRDs host SMBHs with higher masses, we split the 21 LRDs into two bins using the median stellar mass ($M_{\rm \star} = 1.25\times10^{10} \ \rm{M_{\odot}}$) of the sample. Stacking the 10/11 low-mass and high-mass galaxies did not result in a statistically significant detection in either of the stellar mass bins. The bottom panels of Figure~\ref{fig:stacks} show the stacked images for high-mass and low-mass sub-samples. 

Next, we split the LRD galaxies based on their lensing magnification. Because gravitational lensing boosts the X-ray photons associated with the high-redshift galaxies but does not increase the level of ICM emission, stacking galaxies with higher lensing magnification could result in significant detection. To this end, we split the 21 LRDs based on the median lensing magnification ($\mu = 1.69$) of the sample, and co-add the X-ray photons from the low-magnification and high-magnification samples. Yet again, we do not detect a statistically significant detection for the sub-samples. 

Finally, we co-add the sample of nine galaxies from the spectroscopic follow-up of \citet{greene23}, which galaxies exhibit broad-line $H\alpha$ emission, indicative of AGN activity. For this sub-sample, we obtained a tentative, $2.6\sigma$, detection, shown in the top right panel of Figure~\ref{fig:stacks}. To verify that the signal is not due to background fluctuations, we tested different source and background apertures, which resulted in consistent detections at the $\sim2.5\sigma$ significance level. 

These observed values and the limits on the X-ray luminosity and the SMBH masses are further discussed in Section~\ref{sec:bh_mass} and are collated in Table \ref{tab:limits}.

\begin{figure*}[!htp]
  \begin{center}
    \leavevmode
      \epsfxsize=0.9\textwidth \epsfbox{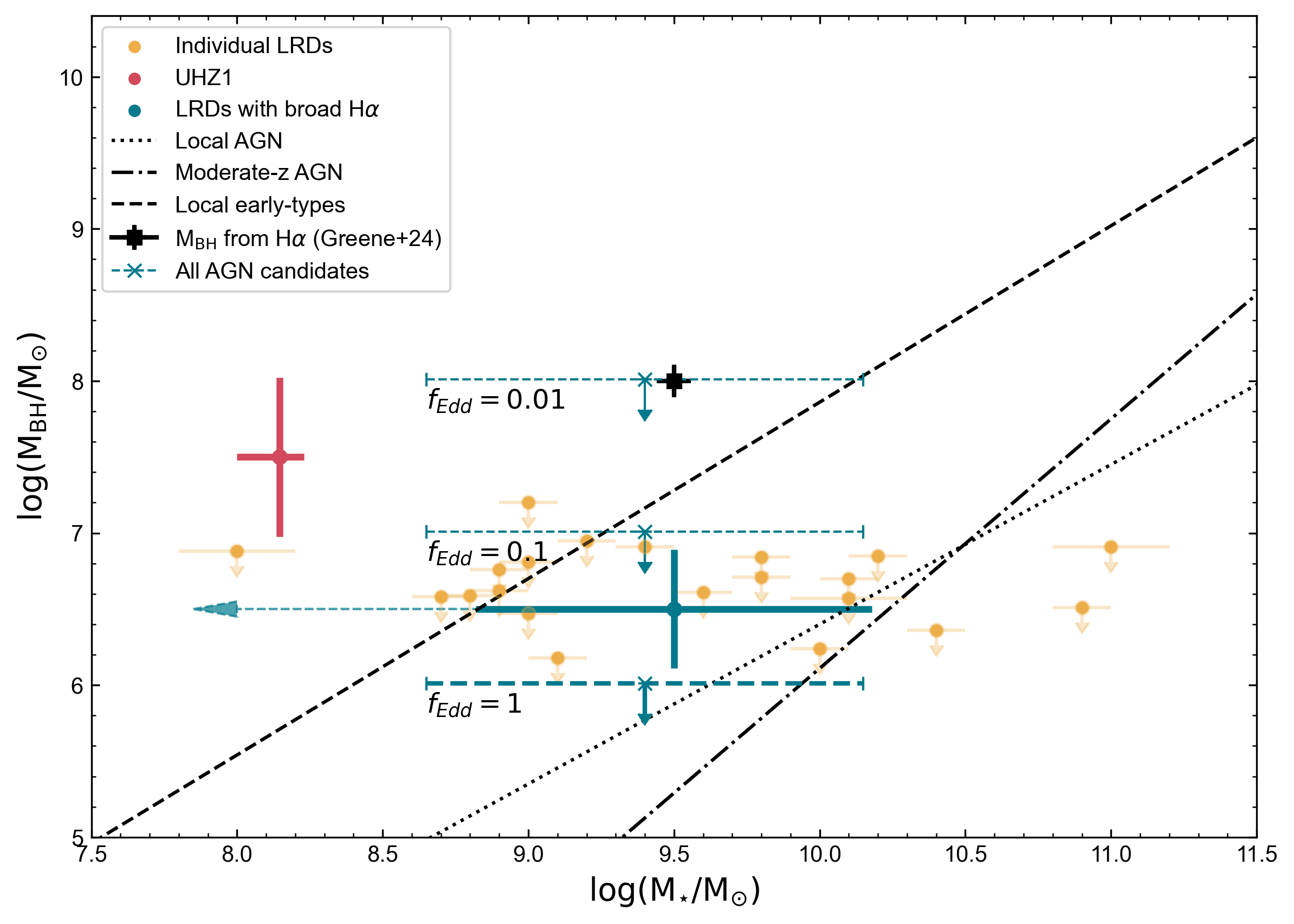}
      \vspace{0.0cm}
      \caption{The relationship between the SMBH mass and the stellar masses of galaxies. The yellow data points represent the upper limits obtained for the individual LRDs. The thick dark green data point corresponds to the upper limit measured for the stacked sample of 21 LRDs assuming Eddington-limited accretion. The uncertainty in stellar mass corresponds to the standard deviation. For reference, we also show the SMBH mass upper limits for the full stacked sample assuming accretion at $10\%$ and $1\%$ of the Eddington limit (green thin dashed). The tentative $2.6\sigma$ detection for the sample of nine LRDs with broad H$\alpha$ line is shown with the solid dark green data point. The arrow pointing left from this data point indicates that the stellar masses used for this plot from \citet{labbe23b} assume galaxy-only fits to calculate the masses. However, galaxy+AGN fits lead to lower AGN masses \citep[e.g.,][]{barro2024} by up to 2 orders of magnitude, therefore this point might shift leftwards, as illustrated by the green dashed arrow. 
      We over-plot the scaling relations observed for local early-type galaxies using the relation observed by \citet{kormendy13} that was scaled following \citet{reines15}, the best-fit relation for local AGN \citep{reines15}, and for AGN up to $z<2.5$ \citep{suh20}. The black data point (square) shows the median SMBH mass inferred from the broad line H$\alpha$ emission line measured by \textit{JWST} \citep{greene23}. We also plot the $z=10.07$ galaxy, UHZ\,1, which hosts an over-massive SMBH \citep{bogdan24}. }
     \label{fig:mbh}
  \end{center}
\end{figure*}

\subsection{SMBH masses}
\label{sec:bh_mass}

In Tables \ref{tab:sample} and \ref{tab:limits}, we provide the observed values and the $2\sigma$ upper limits on the X-ray luminosities and inferred SMBH masses on the individual galaxies and the stacked samples, respectively. Here, we outline the steps that were used to compute these values. 

For the individual galaxies, we computed their $2\sigma$ upper limits using the observed counts in the $2-7$~keV band by applying the Gehrels approximation \citep{gehrels86}. To convert these values to $2-10$~keV band X-ray luminosity, we utilized the exposure maps, assumed a power law model with a slope of $\Gamma=1.9$ and Galactic column density, and used the redshift of each galaxy. We also corrected the observed values by the lensing magnification from the publicly available mass model of Abell~2744 \cite{labbe23b}. From the $2-10$~keV band X-ray luminosity upper limits, we then calculated the bolometric X-ray luminosities using the X-ray-to-bolometric correction \citep{duras20}. We note that in these calculations we do not include the substantial intrinsic scatter associated with the X-ray-to-bolometric correction. The limits on the bolometric luminosities are in the range of $L_{\rm bol}<(0.2-2)\times10^{45} \ \rm{erg \ s^{-1}}$. To derive the upper limits on the SMBH mass, we assumed Eddington-limited accretion. We thereby obtain upper limits on the mass of SMBHs in the range of $M_{\rm BH} < (1.5-15.8) \times 10^{6} \ \rm{M_{\odot}}$. 

For the stacked sample of 21 LRDs and the sub-samples (Section \ref{sec:stacking}), we co-added the X-ray counts associated with the individual galaxies and computed observed value and the $2\sigma$ upper limits on the observed counts following a similar approach as for individual galaxies. The SMBH mass value and limits were calculated using a similar approach and, for the full sample, we place a limit of $M_{\rm BH} < 10^6 \ \rm{M_{\odot}}$ assuming Eddington-limited accretion. We note that this upper limit is $\approx1.5-2$ orders of magnitude lower than the SMBH mass detected for the $z\approx10$ sources, UHZ\,1 and GHZ\,9 \citep{bogdan24,kovacs24}. For the sub-samples, we get fairly similar limits; the SMBHs cannot be more massive than a few times $10^6 \ \rm{M_{\odot}}$ assuming Eddington-limited accretion. Taking the tentative detection at face value for the subset of nine galaxies with broad-line H$\alpha$ emission, we obtain a SMBH mass of $(3.2\pm1.2)\times10^6 \ \rm{M_{\odot}}$.

\section{Discussion}
\label{sec:discussion}

The main finding of our analysis is the non-detection of AGN in individual LRDs and the weak/non-detection of stacked galaxy samples in the \textit{Chandra} X-ray data. Our results are consistent with the X-ray stacking results from \citet{minghao2024}, which also find very weak X-ray emission for a set of 19 LRDs with broad H$\alpha$ emission, and conclude that LRDs might have different properties compared to typical low-redshift Type-1 AGN. The same X-ray weakness was also observed by \citet{maiolino2024}, which studied a sample of 71 JWST-selected AGN and concluded that their X-ray weakness indicates that these AGN are Compton-thick.

Given the deep \textit{Chandra} observations for our sample, the lensing magnification of galaxies, and the stacking approach, we obtain upper limits of the order of $M_{\rm BH} \sim 10^6 \ \rm{M_{\odot}}$ for typical SMBH masses (Tables \ref{tab:sample} and \ref{tab:limits}). Using these highly constraining limits, we place the population of LRDs on the $M_{\rm BH} - M_{\rm \star}$ relation. In Figure \ref{fig:mbh}, we present this relation for the individual and the stacked sample of LRDs. In the plot, we only highlight the limit for all stacked galaxies but the limits on other stacked sub-samples are comparable (Table \ref{tab:limits}). We note that the X-ray upper limits do not show a dependence on the stellar mass of the galaxies, since the limits are primarily driven by the number of counts in the source region, which, in turn, is determined by the depth of the \textit{Chandra} exposure time and the brightness of the ICM of Abell~2744 at that spatial location.  

Comparing the SMBH mass upper limits with the scaling relation for local early-type galaxies \citep{kormendy13}, we find that all but one upper limit lies either below the relations or is consistent within the scatter. While the $M_{\rm BH} - M_{\rm \star}$ scaling relations for local and moderate redshift ($z\lesssim2.5$) AGN have a somewhat lower normalization \citep{reines15,suh20}, most individual upper limits are consistent with these relations. These constraints already suggest that most LRDs are unlikely to host over-massive SMBHs. Similarly, the upper limit for the stacked sample lies about an order of magnitude below the value determined for local early-type galaxies, and is roughly consistent with scaling relations obtained for AGN. We derive a black hole-to-galaxy stellar mass ratio of only $\lesssim0.04\%$ based on the stacked X-ray upper limits. This conclusion is in stark contrast with that observed for many high-redshift galaxies, which suggests that their SMBHs are over-massive relative to their host galaxies and can reach black hole-to-galaxy stellar mass ratios as high as $\sim10\%-100\%$ \citep[e.g.][]{bogdan24,kovacs24,furtak23,li24}. As a caveat, we note that the upper limits on the SMBH masses were calculated assuming Eddington-limited accretion. If, however, we assume that the AGN accrete at $10\%$ of their Eddington-limit, a value lower than what is inferred by \citet{greene23}, the black hole-to-galaxy stellar mass ratio is then $\lesssim0.4\%$. This value would be comparable to what is obtained from the scaling relations but falls significantly short of the ratios observed for several high-redshift AGN. Thus, our X-ray observations suggest that the population of LRDs do not host over-massive SMBHs. Instead, LRDs may host SMBHs, whose masses are consistent with the scaling relations established for local and moderate redshift AGN or they accrete at a small fraction of their Eddington limit.

An additional caveat is that the stellar mass measurements used in this work are obtained from \citet{labbe23b}, which only provides masses determined from galaxy-only SED fits. However, \cite{labbe23b} also notes that including ALMA data prefers galaxy+AGN template fits rather than galaxy-only SED fits. \citet{akins2023,barro2024,wang2024} find AGN signatures in similar high-redshift red objects, and \citet{kocevski2024} find that for LRDs, the rest-frame optical emission likely originates from AGN, while the rest-frame ultraviolet emission originates from the host galaxy. \citet{barro2024} show that if the AGN component is accounted for in mass measurements, the stellar mass could be $1-2$ orders of magnitude lower. 
If we assume that AGN-dominated template fits will lower stellar masses by $1-2$ orders of magnitude  (illustrated by the dashed leftward arrow in Fig. 3), then our results will be consistent with those of galaxies hosting over-massive black holes.

Surprisingly, the tentative $2.6\sigma$ X-ray detection of LRDs with broad-line H$\alpha$ emission contradicts the NIRSpec spectroscopic results obtained by \citet{greene23}. Indeed, the weak X-ray detection implies a SMBH mass of $3.2\times10^6 \ \rm{M_{\odot}}$, which is 1.5 orders of magnitude lower than the median SMBH mass ($10^8 \ \rm{M_{\odot}}$) inferred based on the \textit{JWST} data. The difference between our findings and that of \citet{greene23} is that we assume Eddington-limited accretion, while they infer the mean Eddington limit as $\sim22\%$. Assuming their Eddington limit for our sample, our SMBH mass upper limit would increase to  $\approx1.6\times10^7 \ \rm{M_{\odot}}$, which is still $\approx6$ times lower than the estimate obtained by \citet{greene23}. To reconcile the SMBH masses between the \textit{JWST} near-infrared and the Chandra X-ray observations, an Eddington rate of $\approx3\%$ would be required, which, however, would be inconsistent with the \textit{JWST} measurements. The lack of X-ray emission could also be explained if the accretion disks of the AGN were heavily obscured. However, to hide the $2-7$~keV band X-ray photons observed by \textit{Chandra}, extremely high $N_{\rm H} \gtrsim 10^{25} \ \rm{cm^{-2}}$ columns would be required. This high required column density is inconsistent with those inferred from \textit{JWST} data, which estimate columns of a few times $10^{21} \ \rm{cm^{-2}}$. Alternatively, the difference between the X-ray and near-infrared measurements may stem from systematic uncertainties associated with the determination of SMBH masses from the NIRSpec data. Specifically, \citet{greene23} employed scaling relations between the broad-line region size and the luminosity of the H$\alpha$ lines established for local galaxies \citep{greene05,reines13}. However, this relation may not be applicable for high-redshift, high-accretion rate sources, and in these galaxies, the SMBH mass may therefore be significantly overestimated \citep{du15,linzer22}. Recently, \citet{bertemes24} investigated virial black hole mass estimators and found that numerous factors, such as measurement uncertainties, spectral blending, non-virial conditions, and dust obscuration, can lead to orders of magnitude variations in the inferred SMBH mass. Thus, the contradiction between X-ray and near-infrared data hints that the virial relations to infer SMBH masses may not applicable for high-redshift AGN. 

The discovery of SMBHs in the high-redshift ($z\sim10$) universe suggests that some of these SMBHs likely originated from heavy seeds, i.e.\ from the direct collapse of massive gas clouds \citep{natarajan24}. However, LRDs reside at lower redshifts and their mean inferred SMBH mass does not exceed $\lesssim10^6 \ \rm{M_{\odot}}$. Therefore, when placing these galaxies on BH growth tracks \citep[e.g.][]{kovacs24}, we conclude that SMBHs of LRDs could have originated from light seeds, i.e.\ from the collapse of the first generation of stars, with initial masses of $10-100 \ \rm{M_{\odot}}$. Our results support the earlier theoretical suggestion that multiple seeding mechanisms likely operate simultaneously in the early universe and that offsets from the local BH mass stellar mass relation at higher redshifts may hold important clues to the nature of early BH seeds \cite{natarajan12}. Models tracing the assembly history of black holes over cosmic time, in which seeding ceases by $z \sim 10$ typically report that even then some seeding signatures may persist to late times \cite{RicartePN18}. While LRDs appear to be consistent with originating from low-mass seeds if seed formation does not cease and continues to later cosmic epochs, as proposed by \citet{Natarajan21}, then they could also have originated from late-forming heavy seeds and we could be witnessing extended seeding. At the highest redshifts $z \sim 9-12$, massive seeding is predicted to result in the production of a transient class of galaxies, the over-massive black hole galaxies (OBGs) that are offset in the BH mass - stellar mass scaling relation compared to the local one \cite{Agarwal+13} reflecting initial conditions. At later times, if seeding persists, then where BHs end up on the scaling relations may well be determined primarily once again by seeding rather than their accretion history. The prospects for addressing these important open questions look bright with more detailed follow-up studies of the extended environment of LRDs.

\smallskip

\begin{small}
\noindent
\textit{Acknowledgements.}
We thank the referee for their constructive report. We thank Andy Goulding for the helpful discussions. T.T.A. acknowledges support from ADAP grant 80NSSC24K0692. \'A.B. acknowledges support from the Smithsonian Institution and the Chandra Project through NASA contract NAS8-03060. O.E.K is supported by the GA\v{C}R EXPRO grant No. 21-13491X. P.N. acknowledges support from the Gordon and Betty Moore Foundation and the John Templeton Foundation that fund the Black Hole Initiative (BHI) at Harvard University where she serves as one of the PIs.
This paper employs a list of Chandra datasets, obtained by the Chandra X-ray Observatory, contained in the Chandra Data Collection (CDC) 248~\href{https://doi.org/10.25574/cdc.248}{doi:10.25574/cdc.248}.

\end{small}

\bibliographystyle{aasjournal}
\bibliography{paper1.bib}

\begin{appendix}

\section{Appendix A}

\begin{figure}[h]
    \centering
    \includegraphics[width=0.8\textwidth]{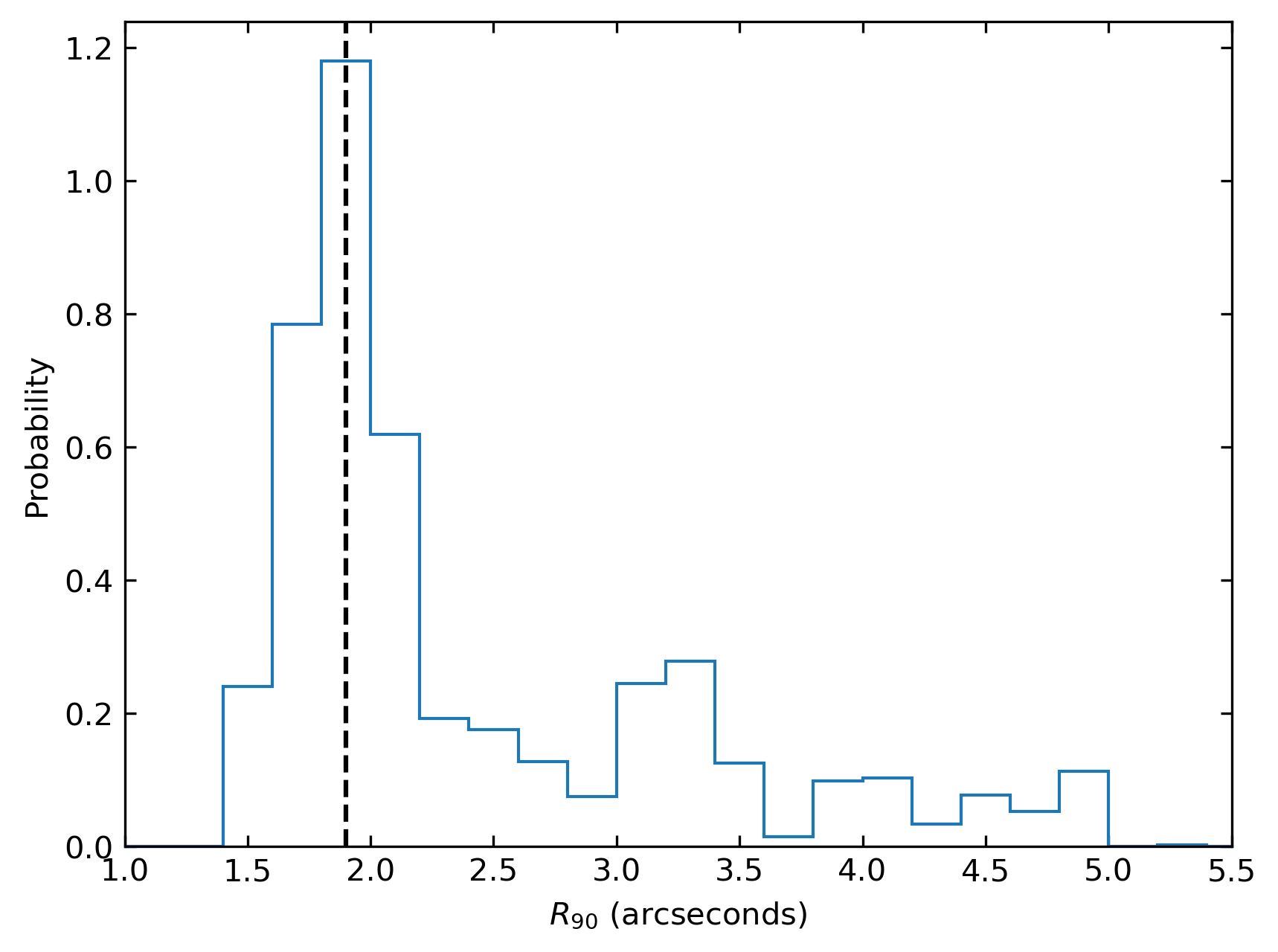}
    \caption{\bf The distribution of 90\% energy encircling radius ($R_{\rm 90}$) for the 21 AGN candidates presented in \citet{labbe23b} in the \textit{Chandra} observations of Abell~2744. This radius varies as a function of distance from \textit{Chandra} ACIS-I pointing center. The median $R_{\rm 90}$ for all source-observation pairs for these sources is $1.9\arcsec$ (shown with black dashed line).}
    \label{fig:r90_histogram}
\end{figure}

For our stacking analysis, photons are extracted by taking into account the $R_{\rm 90}$ value for each source-observation pair. However, in the simplified representation in the co-added image shown in Figure~\ref{fig:stacks}, the source aperture size ($1.9\arcsec$) represents the median $R_{\rm 90}$ for all source-observation pairs for our main sample, the distribution of which is shown in Figure~\ref{fig:r90_histogram}. \textit{Chandra} ACIS-I PSF is described in detail in \href{https://cxc.harvard.edu/ciao/download/doc/psflibman.1.1.pdf}{Chandra PSF library}.

\section{Appendix B}

The analyzed set of \textit{Chandra} ACIS-I observations are listed in \ref{tab:data}.

\begin{table}[!t]
\begin{center}
\caption{\textcolor{black}{Analyzed \textit{Chandra} ACIS-I observations of Abell\,2744.}}
\begin{minipage}{18cm}
\renewcommand{\arraystretch}{1.3}
\centering
\begin{tabular}{ cccc|cccc}
\hline
Observation ID & $t_{\rm exp}^{\rm clean}\ \rm{(ks)}$ & Detector & Observation Date & Observation ID & $t_{\rm exp}^{\rm clean} \ \rm{(ks)}$ & Detector & Observation Date \\
\hline
7915 	 & 18.11 	 & ACIS-I 	 & 2006-11-08 	 &	 25278 	 & 9.78 	 & ACIS-I 	 & 2022-12-02 	 \\
8477 	 & 44.63 	 & ACIS-I 	 & 2007-06-10 	 &	 27575 	 & 19.65 	 & ACIS-I 	 & 2022-12-02 	 \\
8557 	 & 27.30 	 & ACIS-I 	 & 2007-06-14 	 &	 25936 	 & 12.92 	 & ACIS-I 	 & 2023-01-26 	 \\
7712 	 & 8.07 	 & ACIS-I 	 & 2007-09-10 	 &	 27678 	 & 12.42 	 & ACIS-I 	 & 2023-01-27 	 \\
26280 	 & 11.39 	 & ACIS-I 	 & 2022-01-18 	 &	 25939 	 & 14.32 	 & ACIS-I 	 & 2023-01-28 	 \\
25912 	 & 15.10 	 & ACIS-I 	 & 2022-04-18 	 &	 27679 	 & 11.93 	 & ACIS-I 	 & 2023-01-28 	 \\
25911 	 & 16.59 	 & ACIS-I 	 & 2022-04-19 	 &	 27680 	 & 13.21 	 & ACIS-I 	 & 2023-01-28 	 \\
25934 	 & 18.96 	 & ACIS-I 	 & 2022-04-21 	 &	 27681 	 & 9.78 	 & ACIS-I 	 & 2023-01-29 	 \\
25931 	 & 14.55 	 & ACIS-I 	 & 2022-04-23 	 &	 25909 	 & 19.33 	 & ACIS-I 	 & 2023-05-24 	 \\
25954 	 & 13.39 	 & ACIS-I 	 & 2022-04-24 	 &	 27856 	 & 15.88 	 & ACIS-I 	 & 2023-05-25 	 \\
25928 	 & 15.36 	 & ACIS-I 	 & 2022-05-03 	 &	 27857 	 & 12.92 	 & ACIS-I 	 & 2023-05-26 	 \\
25942 	 & 15.18 	 & ACIS-I 	 & 2022-05-04 	 &	 27563 	 & 11.70 	 & ACIS-I 	 & 2023-06-08 	 \\
25958 	 & 11.81 	 & ACIS-I 	 & 2022-05-04 	 &	 25941 	 & 32.65 	 & ACIS-I 	 & 2023-06-09 	 \\
25971 	 & 12.62 	 & ACIS-I 	 & 2022-05-04 	 &	 27896 	 & 13.73 	 & ACIS-I 	 & 2023-06-10 	 \\
25932 	 & 14.08 	 & ACIS-I 	 & 2022-05-05 	 &	 25917 	 & 35.62 	 & ACIS-I 	 & 2023-06-22 	 \\
25972 	 & 31.14 	 & ACIS-I 	 & 2022-05-18 	 &	 25950 	 & 29.69 	 & ACIS-I 	 & 2023-06-30 	 \\
25970 	 & 23.99 	 & ACIS-I 	 & 2022-06-12 	 &	 25946 	 & 29.69 	 & ACIS-I 	 & 2023-07-01 	 \\
25919 	 & 25.03 	 & ACIS-I 	 & 2022-06-13 	 &	 25965 	 & 35.63 	 & ACIS-I 	 & 2023-07-07 	 \\
25920 	 & 29.67 	 & ACIS-I 	 & 2022-06-13 	 &	 25960 	 & 24.76 	 & ACIS-I 	 & 2023-07-08 	 \\
25922 	 & 31.11 	 & ACIS-I 	 & 2022-06-14 	 &	 25926 	 & 61.18 	 & ACIS-I 	 & 2023-07-12 	 \\
25968 	 & 26.92 	 & ACIS-I 	 & 2022-07-12 	 &	 25955 	 & 43.42 	 & ACIS-I 	 & 2023-07-20 	 \\
25967 	 & 33.12 	 & ACIS-I 	 & 2022-08-01 	 &	 25921 	 & 16.87 	 & ACIS-I 	 & 2023-08-04 	 \\
25929 	 & 26.42 	 & ACIS-I 	 & 2022-08-26 	 &	 25959 	 & 15.39 	 & ACIS-I 	 & 2023-08-05 	 \\
25925 	 & 23.41 	 & ACIS-I 	 & 2022-09-02 	 &	 27974 	 & 28.71 	 & ACIS-I 	 & 2023-08-05 	 \\
25956 	 & 13.90 	 & ACIS-I 	 & 2022-09-02 	 &	 25940 	 & 27.72 	 & ACIS-I 	 & 2023-08-10 	 \\
25913 	 & 19.64 	 & ACIS-I 	 & 2022-09-03 	 &	 25966 	 & 18.84 	 & ACIS-I 	 & 2023-08-13 	 \\
25915 	 & 20.31 	 & ACIS-I 	 & 2022-09-03 	 &	 28370 	 & 20.73 	 & ACIS-I 	 & 2023-08-13 	 \\
25923 	 & 10.64 	 & ACIS-I 	 & 2022-09-04 	 &	 25933 	 & 23.88 	 & ACIS-I 	 & 2023-08-15 	 \\
25279 	 & 23.69 	 & ACIS-I 	 & 2022-09-06 	 &	 28483 	 & 20.22 	 & ACIS-I 	 & 2023-08-19 	 \\
25924 	 & 21.54 	 & ACIS-I 	 & 2022-09-07 	 &	 25935 	 & 24.08 	 & ACIS-I 	 & 2023-08-20 	 \\
25944 	 & 20.85 	 & ACIS-I 	 & 2022-09-08 	 &	 27780 	 & 14.90 	 & ACIS-I 	 & 2023-08-21 	 \\
25957 	 & 21.29 	 & ACIS-I 	 & 2022-09-08 	 &	 25943 	 & 16.69 	 & ACIS-I 	 & 2023-08-31 	 \\
27347 	 & 21.45 	 & ACIS-I 	 & 2022-09-09 	 &	 28872 	 & 13.09 	 & ACIS-I 	 & 2023-09-01 	 \\
25918 	 & 20.38 	 & ACIS-I 	 & 2022-09-13 	 &	 25916 	 & 22.20 	 & ACIS-I 	 & 2023-09-03 	 \\
25953 	 & 24.24 	 & ACIS-I 	 & 2022-09-17 	 &	 25964 	 & 20.32 	 & ACIS-I 	 & 2023-09-05 	 \\
25908 	 & 21.83 	 & ACIS-I 	 & 2022-09-23 	 &	 25961 	 & 18.84 	 & ACIS-I 	 & 2023-09-09 	 \\
27449 	 & 9.78 	 & ACIS-I 	 & 2022-09-24 	 &	 28886 	 & 9.96 	 & ACIS-I 	 & 2023-09-10 	 \\
25910 	 & 19.31 	 & ACIS-I 	 & 2022-09-25 	 &	 28887 	 & 19.85 	 & ACIS-I 	 & 2023-09-10 	 \\
27450 	 & 9.78 	 & ACIS-I 	 & 2022-09-26 	 &	 25962 	 & 21.81 	 & ACIS-I 	 & 2023-09-11 	 \\
25945 	 & 16.77 	 & ACIS-I 	 & 2022-09-27 	 &	 25927 	 & 20.53 	 & ACIS-I 	 & 2023-09-16 	 \\
25948 	 & 26.80 	 & ACIS-I 	 & 2022-09-30 	 &	 25947 	 & 14.90 	 & ACIS-I 	 & 2023-09-24 	 \\
25969 	 & 26.92 	 & ACIS-I 	 & 2022-10-09 	 &	 28920 	 & 15.28 	 & ACIS-I 	 & 2023-09-25 	 \\
25914 	 & 27.77 	 & ACIS-I 	 & 2022-10-15 	 &	 25952 	 & 10.84 	 & ACIS-I 	 & 2023-09-27 	 \\
25907 	 & 36.80 	 & ACIS-I 	 & 2022-11-08 	 &	 28934 	 & 19.83 	 & ACIS-I 	 & 2023-09-29 	 \\
25973 	 & 18.15 	 & ACIS-I 	 & 2022-11-11 	 &	 27739 	 & 21.31 	 & ACIS-I 	 & 2023-10-01 	 \\
25930 	 & 19.16 	 & ACIS-I 	 & 2022-11-15 	 &	 25277 	 & 18.68 	 & ACIS-I 	 & 2023-10-02 	 \\
27556 	 & 24.64 	 & ACIS-I 	 & 2022-11-15 	 &	 28951 	 & 12.90 	 & ACIS-I 	 & 2023-10-05 	 \\
25951 	 & 28.45 	 & ACIS-I 	 & 2022-11-18 	 &	 28952 	 & 13.79 	 & ACIS-I 	 & 2023-10-08 	 \\
25938 	 & 18.12 	 & ACIS-I 	 & 2022-11-26 	 &	 28910 	 & 25.75 	 & ACIS-I 	 & 2023-10-25 	 \\
25963 	 & 37.08 	 & ACIS-I 	 & 2022-11-26 	 &	 25949 	 & 20.82 	 & ACIS-I 	 & 2023-10-27 	 \\
25937 	 & 29.73 	 & ACIS-I 	 & 2022-11-27 	 & \\

 \hline
\label{tab:data}
\end{tabular} 
\end{minipage}
\end{center}
\end{table}

\end{appendix}

\end{document}